\documentclass [11pt,a4paper] {article}

\usepackage[cp1252]{inputenc}
\usepackage{amssymb}
\usepackage{amsmath}
\usepackage{amsfonts,amssymb}
\usepackage[dvips]{graphicx}
\usepackage{bbm}
\usepackage{enumerate}
\usepackage{amsthm}
\usepackage{cancel}

\DeclareMathAlphabet{\mathpzc}{OT1}{pzc}{m}{it}

\def\SmallColSep{\setlength{\arraycolsep}{1pt}}

\begin{document}

\title{Quantum state change in light of changes in valuational entropies}

\author{Arkady Bolotin\footnote{$Email: arkadyv@bgu.ac.il$\vspace{5pt}} \\ \emph{Ben-Gurion University of the Negev, Beersheba (Israel)}}

\maketitle

\begin{abstract}\noindent In the statement \textsf{``The vector is an element of the closed linear subspace of the Hilbert space $\mathcal{H}$''}, the predicate \textsf{``\ldots is an element of \ldots''} might be not only determined, that is, either true or false (depending on whether set membership is applicable or inapplicable to the specified vector and subspace) but also undetermined, that is, neither true nor false. To evaluate the vagueness of set membership among arbitrary vectors and closed linear subspaces of $\mathcal{H}$, the notion of the entropy of the predicate \textsf{``\ldots is an element of \ldots''} is introduced in the present paper. Since each closed linear subspace in $\mathcal{H}$ uniquely represents the atomic proposition $P$ about a quantum system, the entropy of this predicate can also be considered as the valuational entropy that measures the uncertainty about the assignment of truth values to the proposition $P$. As it is demonstrated in the paper, in the Hilbert space $\mathcal{H}$ of the dimension greater than or equal to 2, there always exists a nonempty set $S$ of the closed linear subspaces in $\mathcal{H}$, such that the entropy of the predicate \textsf{``\ldots is an element of \ldots''} on the given vector of $\mathcal{H}$ and all the subspaces of $S$ cannot be zero. This implies the existence of two different processes of the pure quantum state change: the process which yields no changes in the valuational entropies of the propositions (corresponding to the deterministic and reversible evolution) and the process which brings forth changes in the valuational entropies (corresponding to the gain or loss of information in a quantum measurement).\\

\end{abstract}

\section{Introduction}  

\noindent In the standard formulation of quantum mechanics (i.e., one introduced by Dirac \cite{Dirac} and von Neumann \cite{Neumann}), the pure (i.e., definite) quantum state of a physical system changes in accordance with two distinct processes. The process of the first kind is involved in measurement (von Neumann referred to it as ``arbitrary change'') and is discontinuous, non-deterministic, and irreversible. Contrastively, the process of the second kind is governed by the Schr{\"o}dinger equation and is continuous, deterministic, and reversible.\\

\noindent What is more, the process of the first kind cannot be reduced to the process of the second kind, and the relationship between the two constitutes the heart of \emph{the quantum measurement problem} \cite{French}.\\

\noindent However, one may argue here that as long as the existence of two separate processes for changing the quantum state is not required by a property (or properties) of a mathematical formalism that provides a rigorous description of quantum mechanics, the measurement problem can be avoided by merely choosing another interpretation, instead of the standard formulation, that would permit a single process for changing the quantum state.\\

\noindent As an example, one may point out the Many Worlds Interpretation (MWI). In MWI, the only presupposed entity is the Wave Function which evolves in agreement with the Schr{\"o}dinger equation (or its relativistic generalizations). Accordingly, MWI permits only the process of the second kind (see references \cite{Everett, Bell, Saunders, Vaidman}, just to name a few).\\

\noindent Hence, the key question is this: Is the existence of two separate processes for changing the quantum state optional? More precisely, is there any basis in a mathematical formalism of quantum mechanics for having two separate processes by which the pure quantum state can change?\\

\noindent The purpose of the present paper is to give the answer to this question in the affirmative.\\

\section{The notion of valuational entropy}  

\noindent Let kets such as $|\Psi\rangle$ denote vectors corresponding to pure quantum states of the system in the Hilbert space $\mathcal{H}$ associated with the system, and calligraphic uppercase letters such as $\mathcal{P}$ denote closed linear subspaces in $\mathcal{H}$.\\

\noindent Recall that in accordance with predicate logic, a verb template, which describes a property of objects or a relation among objects, is called \emph{a predicate} \cite{Cunningham}.\\

\noindent For example, let us consider the statement \textsf{``$|\Psi\rangle$ is an element of $\mathcal{P}\mkern2mu$''}. Clearly, statements such as this can be obtained by substituting appropriate objects for variables $x$ and $y$ in the generic statement \textsf{``$x$ is an element of $y$''}. The template \textsf{``\ldots is an element of \ldots''} is the predicate which describes the relation $\in$ (also called \emph{set membership}) among $x$ and $y$ (equivalently, this predicate can be expressed as \textsf{``\ldots is a member of \ldots''}, \textsf{``\ldots belongs to \ldots''}, \textsf{``\ldots is in \ldots''}, and \textsf{``\ldots lies in \ldots''}). If one adopts the symbol $\mathfrak{P}_{\in}$ as the name for this predicate, then the statement ${x}\in{y}$ can be represented as $\mathfrak{P}_{\in}(x,y)$. Providing $x$ and $y$ are simply placeholders meaning that they can be replaced by any abstract objects, one can use $\mathfrak{P}_{\in}(x,y)$ to denote the predicate \textsf{``\ldots is an element of \ldots''}.\\

\noindent That said, a predicate can also be thought as a propositional function that may return a value that is either true or false depending on the values of its variables. Indeed, in set-builder notation, the predicate $\mathfrak{P}_{\in}(x,y)$ is the rule defining the collection $F$ of all the objects $x$ and $y$ which make the statement ${x}\in{y}$ true:\smallskip

\begin{equation}  
   F 
   =
   \left\{
      x,y
      \mkern3mu
      :
      \mkern3mu
      \mathfrak{P}_{\in}
      (x,y)
   \right\}
   \;\;\;\;   .
\end{equation}
\smallskip

\noindent Consequently, $\mathfrak{P}_{\in}(x,y)$ can be regarded as the image of a tuple $(x,y)$ under the propositional function $\mathfrak{P}_{\in}$ denoted by\smallskip

\begin{equation} \label{IN} 
   \mathfrak{P}_{\in}
   :
   \mathbb{X}
   \times
   \mathbb{Y}
   \to
   \mathbb{B}_{2}
   \;\;\;\;   ,
\end{equation}
\smallskip

\noindent where $\mathbb{X}$ and $\mathbb{Y}$ are the sets of all the objects $x$ and $y$ so that an element of the domain $\mathbb{X} \times \mathbb{Y}$ is a tuple $(x,y)$, while $\mathbb{B}_{2}$ is the set of the truth values, ``true'' and ``false'', renamed to 1 and 0, respectively.\\

\noindent Suppose that the vector $|\Psi\rangle$ is an element of the subspace $\mathcal{P}$ but does not belong to the subspace $\mathcal{Q}$. One says then that the propositional function $\mathfrak{P}_{\in}$ maps the tuples $(|\Psi\rangle,\mathcal{P})$ and $(|\Psi\rangle,\mathcal{Q})$ to the truth values 1 and 0, respectively. This can be abbreviated by $\mathfrak{P}_{\in}(|\Psi\rangle, \mathcal{P}) = 1$ and $\mathfrak{P}_{\in}(|\Psi\rangle, \mathcal{Q}) = 0$. In both cases, the predicate $\mathfrak{P}_{\in}(x,y)$ is determined, that is, the statement ${x}\in{y}$ is either true or false.\\

\noindent Now, consider \emph{borderline cases}, i.e., the penumbra of the predicate $\mathfrak{P}_{\in}(x,y)$. In the borderline cases, the relation $\in$ is neither definitely applicable nor definitely not applicable among the objects $x$ and $y$; interpreted another way, the statement ${x}\in{y}$ is neither true nor false. If $|\Psi\rangle$ and $\mathcal{R}$ are such objects, then this borderline case can be abbreviated by\smallskip

\begin{equation}  
   \mathfrak{P}_{\in}
   \left(
      |\Psi\rangle
      ,
      \mathcal{R}
   \right)
   \notin
   \mathbb{B}_{2}
   \;\;\;\;   .
\end{equation}
\smallskip

\noindent Thus, the propositional function (\ref{IN}) cannot be total but only \emph{partial}. In other words, the function $\mathfrak{P}_{\in}$ is such that some elements of $\mathbb{X} \times \mathbb{Y}$ have no association with $\mathbb{B}_{2}$.\\

\noindent The fact that $\mathfrak{P}_{\in}$ is only a partial function might imply that the predicate $\mathfrak{P}_{\in}(x,y)$ admits \emph{truth value gaps}. In that case, the statement $|\Psi\rangle\in\mathcal{R}$ has no truth value at all, which can be written down in symbols as\smallskip

\begin{equation}  
   \mathfrak{P}_{\in}
   \left(
      |\Psi\rangle
      ,
      \mathcal{R}
   \right)
   =
   0/0
   \;\;\;\;   ,
\end{equation}
\smallskip

\noindent where $0/0$ stands for nonexistent value (i.e., truth value gap).\\

\noindent Alternatively, the fact that the function $\mathfrak{P}_{\in}$ is partial may possibly imply that the predicate $\mathfrak{P}_{\in}(x,y)$ admits \emph{many-valued truth values}. Consequently, the statement $|\Psi\rangle\in\mathcal{R}$ may have any truth value $\tau$ that does not belong to $\mathbb{B}_{2}$. E.g.,\smallskip

\begin{equation}  
   \mathfrak{P}_{\in}
   \left(
      |\Psi\rangle
      ,
      \mathcal{R}
   \right)
   =
   \tau
   \in
   \left(
      0
      ,
      1
   \right)
   \;\;\;\;   .
\end{equation}
\smallskip

\noindent By way of illustration, let us consider the following closed linear subspaces:\smallskip

\begin{equation}  
   \mathcal{P}_{1}
   =
   \left\{
      \!
      \left[
         \begin{array}{r r r r}
            a &
            0 &
            0 &
            0 
         \end{array}
      \right]^{\mathrm{T}}
   \right\}
   \;\;\;\;  ,
\end{equation}
\\[-35pt]

\begin{equation}  
   \mathcal{P}_{2}
   =
   \left\{
      \!
      \left[
         \begin{array}{r r r r}
            0 &
            a &
            0 &
            0 
         \end{array}
      \right]^{\mathrm{T}}
   \right\}
   \;\;\;\;  ,
\end{equation}
\\[-35pt]

\begin{equation}  
   \mathcal{P}_{3}
   =
   \left\{
      \!
      \left[
         \begin{array}{r r r r}
            a &
            a &
            a &
            a 
         \end{array}
      \right]^{\mathrm{T}}
   \right\}
   \;\;\;\;  ,
\end{equation}
\\[-35pt]

\begin{equation}  
   \mathcal{P}_{4}
   =
   \left\{
      \!
      \left[
         \begin{array}{r r r r}
            a &
            a &
            0 &
            0 
         \end{array}
      \right]^{\mathrm{T}}
   \right\}
   \;\;\;\;  ,
\end{equation}
\smallskip

\noindent where $[\cdots\mkern-1mu]^{\mathrm{T}}$  represents transposition operation and $a$ denotes a real number (provided that $a \neq 0$). These subspaces are the subsets of the 4-dimensional Hilbert space $\mathbb{C}^{2 \times 2}$ characterizing the system of two \emph{qubits}, i.e., two-state quantum systems.\\

\noindent Assume that this system is prepared in the state described by the normalized column vector $|\Psi_{1}\rangle$ of $\mathbb{C}^{2 \times 2}$, namely:\smallskip

\begin{equation}  
   |\Psi_{1}\rangle
   =
   \!
   \left[
      \begin{array}{r r r r}
         1 &
         0 &
         0 &
         0 
      \end{array}
   \right]^{\mathrm{T}}
   \;\;\;\;  .
\end{equation}
\smallskip

\noindent Since \emph{all} the components of the scalar product $a|\Psi_{1}\rangle$ match the similar in position components of the element of the subspace $\mathcal{P}_{1}$, the vector $|\Psi_{1}\rangle$ definitely belongs to $\mathcal{P}_{1}$. Hence, the function $\mathfrak{P}_{\in}$ maps the tuple $(|\Psi_{1}\rangle, \mathcal{P}_{1})$ to the value of true, i.e., $\mathfrak{P}_{\in}(|\Psi_{1}\rangle, \mathcal{P}_{1}) = 1$. Oppositely, \emph{all} the components of the scalar product $b|\Psi_{1}\rangle$ match the similar in position components of the element of the subspace $\mathcal{P}_{2}^{\perp}$, the subset of all the vectors in $\mathcal{H}$ orthogonal to the vectors in $\mathcal{P}_{2}$:\smallskip

\begin{equation}  
   \mathcal{P}_{2}^{\perp}
   =
   \left\{
      \!
      \left[
         \begin{array}{r r r r}
            b &
            0 &
            c &
            d 
         \end{array}
      \right]^{\mathrm{T}}
   \right\}
   \;\;\;\;  ,
\end{equation}
\smallskip

\noindent where $b$, $c$ and $d$ are any real numbers (provided that $b\neq0$). This means that the vector $|\Psi_{1}\rangle$ definitely does not belong to $\mathcal{P}_{2}$; consequently, the function $\mathfrak{P}_{\in}$ associates the tuple $(|\Psi_{1}\rangle, \mathcal{P}_{2})$ with the value of false; in symbols, $\mathfrak{P}_{\in}(|\Psi_{1}\rangle, \mathcal{P}_{2}) = 0$.\\

\noindent Furthermore, because \emph{not all} the components of $a|\Psi_{1}\rangle$ match the similar in position components of the element of the subspace $\mathcal{P}_{3}$, the vector $|\Psi_{1}\rangle$ cannot be regarded as an element of $\mathcal{P}_{3}$. But neither can $|\Psi_{1}\rangle$ be regarded as not an element of $\mathcal{P}_{3}$ since \emph{not all} the components of $b|\Psi_{1}\rangle$ match the similar in position components of the element of $\mathcal{P}_{3}^{\perp}$, the set of all the vectors in $\mathcal{H}$ orthogonal to the vectors lying in $\mathcal{P}_{3}$, namely,\smallskip

\begin{equation}  
   \mathcal{P}_{3}^{\perp}
   =
   \left\{
      \!
      \left[
         \begin{array}{r r r r}
            b &
    -b-c-d &
            c &
            d 
         \end{array}
      \right]^{\mathrm{T}}
   \right\}
   \;\;\;\;  .
\end{equation}
\smallskip

\noindent Hence, the relation $\in$ is neither definitely applicable nor definitely inapplicable among $|\Psi_{1}\rangle$ and $\mathcal{P}_{3}$. As a result, the value of the function $\mathfrak{P}_{\in}$ at $(|\Psi_{1}\rangle, \mathcal{P}_{3})$ is neither 1 nor 0, that is, $\mathfrak{P}_{\in}(|\Psi_{1}\rangle, \mathcal{P}_{3}) \notin \mathbb{B}_{2}$.\\

\noindent Likewise, because \emph{not all} the components of $a|\Psi_{1}\rangle$ match the components of the elements of $\mathcal{P}_{4}$, as well as \emph{not all} the components of $-b|\Psi_{1}\rangle$ match the components of the element of $\mathcal{P}_{4}^{\perp}$, the subset of all the vectors in $\mathcal{H}$ orthogonal to the vectors belonging to $\mathcal{P}_{4}$,\smallskip

\begin{equation}  
   \mathcal{P}_{4}^{\perp}
   =
   \left\{
      \!
      \left[
         \begin{array}{r r r r}
           -b &
            b &
            c &
            d 
         \end{array}
      \right]^{\mathrm{T}}
   \right\}
   \;\;\;\;  ,
\end{equation}
\smallskip

\noindent the tuple $(|\Psi_{1}\rangle, \mathcal{P}_{4})$ has no association with the Boolean codomain $\mathbb{B}_{2}$ under the function $\mathfrak{P}_{\in}$, i.e., $\mathfrak{P}_{\in}(|\Psi_{1}\rangle, \mathcal{P}_{4}) \notin \mathbb{B}_{2}$.\\

\noindent To evaluate the vagueness of the relation $\in$ among arbitrary $x$ and $y$, one can bring into play an entropy of the predicate $\mathfrak{P}_{\in}(x,y)$.\\

\noindent Let $\mathcal{P}^{\perp}$ be the subset of all the vectors in the $N$-dimensional Hilbert space $\mathcal{H}$ that are orthogonal to the vectors in the subspace $\mathcal{P}\subseteq\mathcal{H}$, and let $\mathbf{p}$ and $\mathbf{p}^{\perp}\mkern1mu$, the elements of $\mathcal{P}$ and $\mathcal{P}^{\perp}$, respectively, be expressed as the column vectors whose components are $\mathrm{p}_{i}$ and $\mathrm{p}_{i}^{\perp}\mkern1mu$, in that order, where $1 \le i \le N$. Suppose that $|\Psi\rangle$ is the column vector in $\mathcal{H}$, and let $\mathrm{u}_{i}$ denote the components of $|\Psi\rangle$. Consider the sets $M$ and $M^{\perp}$ containing the components of the scalar products $a|\Psi\rangle$ and $b|\Psi\rangle$ which have the counterparts in $\mathbf{p}$ and $\mathbf{p}^{\perp}$, correspondingly, that is,\smallskip

\begin{equation}  
   M
   =
   \left\{
      a\mathrm{u}_{i}
      \mkern1mu
      :
      \mkern3mu
      a\mathrm{u}_{i}
      =
      \mathrm{p}_{i}
   \right\}
   \;\;\;\;  ,
\end{equation}
\\[-35pt]

\begin{equation}  
   M^{\perp}
   =
   \left\{
      b\mathrm{u}_{i}
      \mkern1mu
      :
      \mkern3mu
      b\mathrm{u}_{i}
      =
      \mathrm{p}_{i}^{\perp}
   \right\}
   \;\;\;\;  .
\end{equation}
\smallskip

\noindent Then, the entropy $\mathrm{H}$ of the predicate $\mathfrak{P}_{\in}(x,y)$ on the vector $|\Psi\rangle$ and the subspace $\mathcal{P}$ can be calculated as follows:\smallskip

\begin{equation}  
   \mathrm{H}
   \big(
      \mathfrak{P}_{\in}
      \left(
         |\Psi\rangle
         ,
         \mathcal{P}
      \right)
      \mkern-3mu
   \big)
   =
   \log_{\beta}N
   -
   \frac{
      \max{
      \mkern-3mu
         \left(
            |M|
            ,
            |M^{\perp}|
         \right)
      }
      \log_{\beta}
      \max{
      \mkern-3mu
         \left(
            |M|
            ,
            |M^{\perp}|
         \right)
      }
   }
   {N}
   \;\;\;\;  ,
\end{equation}
\smallskip

\noindent where $|M|$ and $|M^{\perp}|$ stand for the cardinality of the sets $M$ and $M^{\perp}$, while $\beta$ is the base of the logarithm used.\\

\noindent Imagine that  $|M|$ or $|M^{\perp}|$ is equal to $N$. At that case, either all the components of $a|\Psi\rangle$ have the counterparts in the element of $\mathcal{P}$ or all the components of $b|\Psi\rangle$ have the counterparts in the element of $\mathcal{P}^{\perp}$, meaning that the statement $|\Psi\rangle\in\mathcal{P}$ is either true or false. Accordingly, the function $\mathfrak{P}_{\in}$ at $(|\Psi\rangle,\mathcal{P})$ returns a value that is either 1 or 0, and thus the entropy $\mathrm{H}(\mathfrak{P}_{\in}(|\Psi\rangle,\mathcal{P}))$ is zero. The uncertainty over whether $|\Psi\rangle$ belongs to $\mathcal{P}$ or whether $|\Psi\rangle$ does not belong to $\mathcal{P}$ emerges when both $|M|$ and $|M^{\perp}|$ are less than $N$. This uncertainty is quantified in the greater than zero entropy $\mathrm{H}(\mathfrak{P}_{\in}(|\Psi\rangle,\mathcal{P}))$.\\

\noindent As an example, for the tuple $(|\Psi_{1}\rangle,\mathcal{P}_{3})$, the cardinalities of the sets $M$ and $M^{\perp}$ are 1 and 3; therefore, it is uncertain whether or not $|\Psi_{1}\rangle$ belongs to $\mathcal{P}_{3}$. The entropy of the predicate $\mathfrak{P}_{\in}(x,y)$ on $|\Psi_{1}\rangle$ and $\mathcal{P}_{3}$ gives quantity to this uncertainty, namely, $\mathrm{H}(\mathfrak{P}_{\in}(|\Psi_{1}\rangle,\mathcal{P}_{3})) = \log_{\beta}4 - \frac{3}{4}\log_{\beta}3$.\\

\noindent In this manner, the entropy $\mathrm{H}$ of the predicate $\mathfrak{P}_{\in}(x,y)$ on $|\Psi\rangle$ and $\mathcal{P}$ is zero when this predicate is determined. The entropy $\mathrm{H}(\mathfrak{P}_{\in}(|\Psi\rangle,\mathcal{P}))$ is greater than zero when the relation $\in$ is neither definitely applicable nor definitely inapplicable among $|\Psi\rangle$ and $\mathcal{P}$. This can be summarized as follows:\smallskip

\begin{equation} \label{ENT} 
   \mathrm{H}
   \big(
      \mathfrak{P}_{\in}
      \left(
         |\Psi\rangle
         ,
         \mathcal{P}
      \right)
   \big)
   =
   \left\{
      \begingroup\SmallColSep
      \begin{array}{r l}
         0
         \mkern5mu
         ,
         &
         \mkern8mu
         |\Psi\rangle
         \in
         \mathcal{P}
         \mkern5mu
         \text{is either true or false}
         \\
         \\[-13pt]
         h
         \in
         \left(
            0
            ,
            \mathrm{H}_{\text{max}}
            \right.
         \left.
         \mkern-6mu
         \right]
         ,
         &
         \mkern8mu
         |\Psi\rangle
         \in
         \mathcal{P}
         \mkern5mu
         \text{is neither true nor false}
      \end{array}
      \endgroup   
   \right.
   \;\;\;\;  ,
\end{equation}
\smallskip

\noindent where $\mathrm{H}_{\text{max}}$ is the maximum of the entropy $\mathrm{H}(\mathfrak{P}_{\in}(|\Psi\rangle,\mathcal{P}))$.\\

\noindent Given that each closed linear subspace of system's Hilbert space $\mathcal{H}$ uniquely corresponds to the accordant elementary statement (\emph{atomic proposition}) about this system \cite{Birkhoff, Mackey, Redei}, one can say that the predicate $\mathfrak{P}_{\in}(x,y)$ on the state $|\Psi\rangle$ and the subspace $\mathcal{P}$ determines the truth value of the proposition $P$ that is represented by $\mathcal{P}$, namely,\smallskip

\begin{equation} \label{VAL} 
   \mathfrak{P}_{\in}
   \left(
      |\Psi\rangle
      ,
      \mathcal{P}
   \right)
   =
   {[\mkern-3.3mu[
      P
   ]\mkern-3.3mu]}_v
   \;\;\;\;  ,
\end{equation}
\smallskip

\noindent where the double-bracket notation denotes \emph{a valuation} \cite{Dalen, Dunn}, that is, a mapping from the set of atomic propositions, symbolized by $\mathbb{P}$, to the Boolean codomain $\mathbb{B}_{2}$, i.e.,\smallskip

\begin{equation}  
   v
   \mkern-3.3mu
   :
   \mkern2mu
   \mathbb{P}
   \to
   \mathbb{B}_{2}
   \;\;\;\;  ,
\end{equation}
\smallskip

\noindent such that $v(P) = {[\mkern-3.3mu[P]\mkern-3.3mu]}_v$.\\ 

\noindent Consequently, the entropy of the predicate $\mathfrak{P}_{\in}(x,y)$ on $|\Psi\rangle$ and $\mathcal{P}$ (evaluating the vagueness of the relation $\in$ among $|\Psi\rangle$ and $\mathcal{P}$) can also be called \emph{valuational entropy} of the proposition $P$, i.e., $\mathrm{H}({[\mkern-3.3mu[P]\mkern-3.3mu]}_v)$. This entropy measures the uncertainty about the assignment of truth values to proposition $P$, with the result that $\mathrm{H}({[\mkern-3.3mu[P]\mkern-3.3mu]}_v)$ takes on zero when $P$ has either value in $\mathbb{B}_{2}$, and $\mathrm{H}({[\mkern-3.3mu[P]\mkern-3.3mu]}_v)$ is greater than zero when $P$ is neither true nor false, i.e., has no association with $\mathbb{B}_{2}$.\\

\noindent Let $\mathbf{X}$ stand for a random binary variable taking on values ${x}\in\{0,1\}$, explicitly,\smallskip

\begin{equation}  
   \mathbf{X}(\omega)
   =
   \left\{
      \begingroup\SmallColSep
      \begin{array}{r l}
         1
         ,
         &
         \mkern15mu
         \omega
         =
         1
         \\
         0
         ,
         &
         \mkern15mu
         \omega
         =
         2
      \end{array}
      \endgroup   
   \right.
   \;\;\;\;  ,
\end{equation}
\smallskip

\noindent whose probability mass function is denoted by $\Pr(\mathbf{X}=x)$. Then, \emph{the information entropy} of the variable $\mathbf{X}$ can be calculated using Shannon's formula \cite{Shannon} as follows:\smallskip

\begin{equation}  
   \mathrm{H}
   \left(
      \mathbf{X}
   \right)
   =
   -
   \sum_{x = 0}^1
      \Pr
      \left(
         \mathbf{X}
         =
         x
      \right)
      \log_{\gamma}
      \Pr
      \left(
         \mathbf{X}
         =
         x
      \right)
   \;\;\;\;  ,
\end{equation}
\smallskip

\noindent where the base $\gamma$ determines the units of the information entropy. In case of either value of $\mathbf{X}$ is impossible, this entropy will be the lowest one, i.e., $\mathrm{H}(\mathbf{X}) = 0$, as there will be no uncertainty concerning the value of the variable $\mathbf{X}$. The entropy $\mathrm{H}(\mathbf{X})$ is greater than zero when it is impossible to tell with certainty the value of $\mathbf{X}$.\\

\noindent Following the assumption that the truth values can be interpreted as values conveying  information about propositions \cite{Shramko}, one can present the connection between the valuational entropy $\mathrm{H}({[\mkern-3.3mu[P]\mkern-3.3mu]}_v)$ and the information entropy $\mathrm{H}(\mathbf{X})$ as follows:\smallskip

\begin{equation}  
   \begin{array}{r l}
      \begin{array}{r}
         {[\mkern-3.3mu[P]\mkern-3.3mu]}_v
         \in
         \mathbb{B}_{2}
         \\
         \\[-8.5pt]
         \Pr(\mathbf{X}=x)
         \in
         \{0,1\}
      \end{array}
      :
      &
      \mkern10mu
      \mathrm{H}({[\mkern-3.3mu[P]\mkern-3.3mu]}_v)
      =
      \mathrm{H}(\mathbf{X})
      =
      0
      \\
      \\[-2.5pt]
      \begin{array}{r}
         {[\mkern-3.3mu[P]\mkern-3.3mu]}_v
         \notin
         \mathbb{B}_{2}
         \\
         \\[-8.5pt]
         \Pr(\mathbf{X}=x)
         \in
         (0,1)
      \end{array}
      :
      &
      \begin{array}{r}
         \mathrm{H}({[\mkern-3.3mu[P]\mkern-3.3mu]}_v)
         >
         0
         \\
         \\[-8.5pt]
         \mathrm{H}(\mathbf{X})
         >
         0
      \end{array}
   \end{array}
    \;\;\;\;  .
\end{equation}
\smallskip

\noindent One might infer from here that the truth value ${[\mkern-3.3mu[P]\mkern-3.3mu]}_v$ is equal to the probability $\Pr(\mathbf{X}=x)$. In line with such an inference, the probability $\Pr(\mathbf{X}=x)$ represents \emph{the degree of truth} to which the vector $|\Psi\rangle$ belongs to the subspace $\mathcal{P}$ (see works \cite{Pykacz10, Pykacz11, Pykacz15, Pykacz15b} where this point of view is developed in detail).\\

\section{Two processes of quantum state changes}  

\noindent Imagine the transformation of the truth values of the atomic propositions about the quantum system over time. Assume that this transformation can be presented by some function $f$ from $\mathbb{B}_{2}$ to $\mathbb{B}_{2}$ defined by\smallskip

\begin{equation}  
   \begin{array}{r l}
      f
      :
      &
      \mathbb{B}_{2}
      \to
      \mathbb{B}_{2}
      \\
      \\[-8pt]
      \;
      &
      {[\mkern-3.3mu[P_{\text{past}}]\mkern-3.3mu]}_v
      \mapsto
      {[\mkern-3.3mu[P_{\text{present}}]\mkern-3.3mu]}_v
   \end{array}
    \;\;\;\;  ,
\end{equation}
\smallskip

\noindent where $P_{\text{past}}$ and $P_{\text{present}}$ denote the past-tense and present-tense forms of the atomic proposition $P$, respectively. These tensed forms indicate that $P$ might be true (false) in the past and $P$ may be true (false) at the present, in that order.\\

\noindent For example, take the ``tenseless'' atomic proposition asserting that the spin of the qubit along the $x$-axis is $+\frac{\hbar}{2}$. The past-tense form of this proposition would be \textsf{``In the past, the spin of the qubit along the $x$-axis was $+\frac{\hbar}{2}\mkern3.3mu$''}, whereas its present tense is \textsf{``At this moment in time, the spin of the qubit along the $x$-axis is $+\frac{\hbar}{2}\mkern3.3mu$''} (see \cite{Wolterstorff} for the review on tenses of propositions).\\

\noindent Let us reflect on the situation where the valuational entropy of any atomic proposition $P$ about the system does not change over time, namely,\smallskip

\begin{equation}  
   \Delta
   \mathrm{H}
   \mkern-3.3mu
   \left(
      {[\mkern-3.3mu[
         P
      ]\mkern-3.3mu]}_v
   \right)
   =
   \mathrm{H}
   \mkern-3.3mu
   \left(
      {[\mkern-3.3mu[
         P_{\text{present}}
      ]\mkern-3.3mu]}_v
   \right)
   -
   \mathrm{H}
   \mkern-3.3mu
   \left(
      {[\mkern-3.3mu[
         P_{\text{past}}
      ]\mkern-3.3mu]}_v
   \right)
   =
   0
   \;\;\;\;  .
\end{equation}
\smallskip

\noindent The said situation includes the case of the proposition $P$ that had either value in $\mathbb{B}_{2}$ in the past and continues to have either value in $\mathbb{B}_{2}$ now. In that case, each element of $\mathbb{B}_{2}$ must be paired with itself or (and) another element of $\mathbb{B}_{2}$. But as long as the principle of bivalence holds (which states that a proposition cannot be both true and false), each element of $\mathbb{B}_{2}$ may be paired with either itself or another element of $\mathbb{B}_{2}$. Therefore, $f$ must be a permutation of truth values in $\mathbb{B}_{2}$.\\

\noindent Rewriting the formulas (\ref{IN}) and (\ref{VAL}) such that\smallskip

\begin{equation}  
   \begin{array}{r l}
      \mathfrak{P}_{\in}
      :
      &
      \mathbb{X}
      \times
      \mathbb{Y}
      \to
      \mathbb{B}_{2}
      \\
      \\[-8pt]
      \;
      &
      \left(
         |\Psi\rangle
         ,
         \mathcal{P}
      \right)
      \mapsto
      {[\mkern-3.3mu[P]\mkern-3.3mu]}_v
   \end{array}
    \;\;\;\;  ,
\end{equation}
\smallskip

\noindent one can say then that $f$ is a bijective function from the domain $\mathbb{X}\times\mathbb{Y}$ to itself, that is, a function for which every tuple $(x,y)$ occurs exactly once as an image value:\smallskip

\begin{equation} \label{FUNC} 
   f
   \mkern-3.3mu
   :
   \mathbb{X}
   \times
   \mathbb{Y}
   \to
   \mkern3mu
   \mathbb{X}
   \times
   \mathbb{Y}
    \;\;\;\;  .
\end{equation}
\smallskip

\noindent Since the first argument, $x$, represents the quantum state $|\Psi\rangle$ at different points in time and the second argument, $y$, is fixed to the particular closed linear subspace $\mathcal{P}$ representing the proposition $P$, this produces the partially applied function\smallskip

\begin{equation}  
   |\Psi_{\text{past}}\rangle
   \to
   \mkern3mu
   |\Psi_{\text{present}}\rangle
    \;\;\;\;  ,
\end{equation}
\smallskip

\noindent meaning that the current state of the system $|\Psi_{\text{present}}\rangle$ has evolved from its initial state $|\Psi_{\text{past}}\rangle$ by a deterministic and reversible evolution.\\

\noindent Let us analyze the situation where the valuational entropies of the propositions change over time. Suppose that the valuational entropy of the proposition $P$ goes up from zero or goes down to zero, i.e.,\smallskip

\begin{equation}  
   \Delta
   \mathrm{H}
   \mkern-3.3mu
   \left(
      {[\mkern-3.3mu[
         P
      ]\mkern-3.3mu]}_v
   \right)
   =
   \left\{
      \begingroup\SmallColSep
      \begin{array}{r r}
         \mathrm{H}
         \mkern-3.3mu
         \left(
            {[\mkern-3.3mu[
               P_{\text{present}}
            ]\mkern-3.3mu]}_v
         \right)
         \neq
         0
         ,
         &
         \mkern15mu
         \mathrm{H}
         \mkern-3.3mu
         \left(
            {[\mkern-3.3mu[
               P_{\text{past}}
            ]\mkern-3.3mu]}_v
         \right)
         =
         0
         \\[7pt]
         -
         \mathrm{H}
         \mkern-3.3mu
         \left(
            {[\mkern-3.3mu[
               P_{\text{past}}
            ]\mkern-3.3mu]}_v
         \right)
         \neq
         0
         ,
         &
         \mkern15mu
         \mathrm{H}
         \mkern-3.3mu
         \left(
            {[\mkern-3.3mu[
               P_{\text{present}}
            ]\mkern-3.3mu]}_v
         \right)
         =
         0
      \end{array}
      \endgroup   
   \right.
   \;\;\;\;  .
\end{equation}
\smallskip

\noindent In accordance with (\ref{ENT}) and (\ref{VAL}), this means that at some point in time, the function value $\mathfrak{P}_{\in}(|\Psi\rangle, \mathcal{P})$ does not define a truth value (belonging to $\mathbb{B}_{2}$) for the proposition $P$. Hence, in this situation, the elements of the set $\mathbb{B}_{2}$ are not coupled, that is, each element of $\mathbb{B}_{2}$ is paired with neither itself nor another element of $\mathbb{B}_{2}$. Therefore, in that case, the function (\ref{FUNC}) cannot be bijective and so reversible. As a result, the state $|\Psi_{\text{past}}\rangle$ cannot be changed into the state $|\Psi_{\text{present}}\rangle$ in a deterministic and reversible manner.\\

\noindent To sum up, if processes of the quantum state change yield no differences in the valuational entropies of the propositions (i.e., cause neither gain nor loss of information about the propositions), then such processes are deterministic and reversible. By contrast, if processes of the quantum state change bring forth changes in the valuational entropies of the propositions (i.e., cause the gain or loss of information about the propositions), then these processes are nondeterministic and nonreversible.\\

\noindent The aforesaid deterministic and reversible processes can be regarded as normal or Schr{\"o}dinger's evolutions. As to the nondeterministic and nonreversible processes, those, which are associated with the minimizing of the uncertainty about the truth value assignment, can be regarded as the ``collapse'' of the quantum state, while those, which are associated with the maximizing of the uncertainty about the truth value assignment, can be regarded as the loss of information in a quantum measurement.\\

\section{Vagueness of the statement ${x}\!\in\!{y}$ is inherent in a Hilbert space}  

\noindent The existence of two distinct processes for changing the pure quantum state stems from the fact that the predicate $\mathfrak{P}_{\in}(x,y)$ can be determined as well as undetermined.\\

\noindent Really, imagine that the predicate $\mathfrak{P}_{\in}(x,y)$ was determined on any arbitrary $|\Psi\rangle$ and $\mathcal{P}$. In that case, according to the formula (\ref{VAL}), each proposition would have a truth value in any pure quantum state, and, thus, the valuational entropy of each proposition would always be zero. Consequently, changing the pure quantum state would bring no changes in the valuational entropies and therefore every process of the quantum state change would be deterministic and reversible (put differently, there would be no place for the negative or positive change in the valuational entropy, that is, for the quantum state ``collapse'' or ``expansion'').\\

\noindent But vagueness of the predicate $\mathfrak{P}_{\in}(x,y)$ on arbitrary $|\Psi\rangle$ and $\mathcal{P}$ is \emph{inherent} in a $N$-dimensional Hilbert space $\mathcal{H}$ with $N \ge 2$. Indeed, consistent with the Kochen-Specker theorem \cite{Kochen}, one can assert that there always exists a nonempty set $S =\{ \mathcal{P},\mathcal{Q},\dots\}$ of closed linear subspaces in $\mathcal{H}$, such that the entropy $\mathrm{H}$ of the predicate $\mathfrak{P}_{\in}(x,y)$ on the given vector $|\Psi\rangle$ in $\mathcal{H}$ and all the subspaces of the set $S$ cannot be zero.\\

\noindent Subsequently, the existence of zero and nonzero entropies of the predicate $\mathfrak{P}_{\in}(x,y)$ on $|\Psi\rangle$ and ${y}\in{S}$ involves zero and nonzero changes in the valuational entropies of propositions, that is, the existence of two distinct processes of the state $|\Psi\rangle$ change.\\

\noindent To be sure, in the case of the two-dimensional Hilbert space $\mathbb{C}^2$, one can find the set\smallskip

\begin{equation}  
   S
   =
   \left\{
      \mathcal{X}_{+}
      ,
      \mathcal{Z}_{+}
      ,
      \mathcal{Z}_{-}
   \right\}
   \;\;\;\;    
\end{equation}
\smallskip

\noindent whose members stand for the closed linear subspaces of $\mathbb{C}^2$, explicitly,\smallskip

\begin{equation}  
   \mathcal{X}_{+}
   =
   \left\{
      \!
      \left[
         \begingroup\SmallColSep
         \begin{array}{r r}
            a &
            a 
         \end{array}
         \endgroup
      \right]^{\mathrm{T}}
   \right\}
   \;\;\;\;  ,
\end{equation}
\\[-35pt]

\begin{equation}  
   \mathcal{Z}_{+}
   =
   \left\{
      \!
      \left[
         \begingroup\SmallColSep
         \begin{array}{r r}
            a &
            0 
         \end{array}
         \endgroup
      \right]^{\mathrm{T}}
   \right\}
   \;\;\;\;  ,
\end{equation}
\\[-35pt]

\begin{equation}  
   \mathcal{Z}_{-}
   =
   \left\{
      \!
      \left[
         \begingroup\SmallColSep
         \begin{array}{r r}
            0 &
            a 
         \end{array}
         \endgroup
      \right]^{\mathrm{T}}
   \right\}
   \;\;\;\;   ,
\end{equation}
\smallskip

\noindent one-to-one representing the atomic propositions \textsf{``The spin of the qubit along the $x$-axis is $+\frac{\hbar}{2}\mkern3.3mu$''}, \textsf{``The spin of the qubit along the $z$-axis is $+\frac{\hbar}{2}\mkern3.3mu$''}, and \textsf{``The spin of the qubit along the $z$-axis is $-\frac{\hbar}{2}\mkern3.3mu$''}, replaced for brevity with the letters $X_{+}$, $Z_{+}$, and $Z_{-}$, respectively. Given the pure quantum state $|\Psi_{z+}\rangle$ in $\mathbb{C}^2$\smallskip

\begin{equation}  
   |\Psi_{z+}\rangle
   =
   \!
   \left[
      \begingroup\SmallColSep
      \begin{array}{r r}
         1 &
         0 
      \end{array}
      \endgroup
   \right]^{\mathrm{T}}
   \;\;\;\;  ,
\end{equation}
\smallskip

\noindent one finds\smallskip

\begin{equation}  
   \mathrm{H}
   \big(
      \mathfrak{P}_{\in}
      \left(      
         |\Psi_{z+}\rangle
         ,
         \mathcal{X}_{+}
      \right)
      \mkern-3mu
   \big)
   =
   \log_{\beta}2
   \;\;\;\;  ,
\end{equation}
\\[-35pt]

\begin{equation}  
   \mathrm{H}
   \big(
      \mathfrak{P}_{\in}
      \left(      
         |\Psi_{z+}\rangle
         ,
         \mathcal{Z}_{\pm}
      \right)
      \mkern-3mu
   \big)
   =
   0
   \;\;\;\;   ,
\end{equation}
\smallskip

\noindent where $\mathcal{Z}_{\pm}$ should be replaced by either $\mathcal{Z}_{+}$ or $\mathcal{Z}_{-}$.\\

\noindent On the other hand, given the normalized vectors $|\Psi_{z-}\rangle$ and $|\Psi_{x+}\rangle$ in $\mathbb{C}^2$, namely,\smallskip

\begin{equation}  
   |\Psi_{z-}\rangle
   =
   \!
   \left[
      \begingroup\SmallColSep
      \begin{array}{r r}
         0 &
         1 
      \end{array}
      \endgroup
   \right]^{\mathrm{T}}
   \;\;\;\;  ,
\end{equation}
\\[-35pt]

\begin{equation}  
   |\Psi_{x+}\rangle
   =
   \!
   \frac{1}{\sqrt{2}}
   \left[
      \begingroup\SmallColSep
      \begin{array}{r r}
         1 &
         1 
      \end{array}
      \endgroup
   \right]^{\mathrm{T}}
   \;\;\;\;  ,
\end{equation}
\smallskip

\noindent one finds\smallskip

\begin{equation}  
   \mathrm{H}
   \big(
      \mathfrak{P}_{\in}
      \left(      
         |\Psi_{z-}\rangle
         ,
         \mathcal{X}_{+}
      \right)
      \mkern-3mu
   \big)
   =
   \log_{\beta}2
   \;\;\;\;  ,
\end{equation}
\\[-35pt]

\begin{equation}  
   \mathrm{H}
   \big(
      \mathfrak{P}_{\in}
      \left(      
         |\Psi_{z-}\rangle
         ,
         \mathcal{Z}_{\pm}
      \right)
      \mkern-3mu
   \big)
   =
   0
   \;\;\;\;  ,
\end{equation}
\smallskip

\noindent in addition to\smallskip

\begin{equation}  
   \mathrm{H}
   \big(
      \mathfrak{P}_{\in}
      \left(      
         |\Psi_{x+}\rangle
         ,
         \mathcal{X}_{+}
      \right)
      \mkern-3mu
   \big)
   =
   0
   \;\;\;\;  ,
\end{equation}
\\[-35pt]

\begin{equation}  
   \mathrm{H}
   \big(
      \mathfrak{P}_{\in}
      \left(      
         |\Psi_{x+}\rangle
         ,
         \mathcal{Z}_{\pm}
      \right)
      \mkern-3mu
   \big)
   =
   \log_{\beta}2
   \;\;\;\;  .
\end{equation}
\smallskip

\noindent It follows then that two different processes of the quantum state $|\Psi_{z+}\rangle$ change are possible: The process of the deterministic and reversible kind, $|\Psi_{z+}\rangle \mapsto |\Psi_{z-}\rangle$, corresponds to no changes in the valuational entropies of the propositions $X_{+}$, $Z_{+}$ and $Z_{-}$:\smallskip

\begin{equation}  
   \Delta
   \mathrm{H}
   \left(
      {[\mkern-3.3mu[
         X_{+}
      ]\mkern-3.3mu]}_v
   \right)
   =
   \mathrm{H}
   \big(
      \mathfrak{P}_{\in}
      \left(      
         |\Psi_{z-}\rangle
         ,
         \mathcal{X}_{+}
      \right)
      \mkern-3mu
   \big)
   -
   \mathrm{H}
   \big(
      \mathfrak{P}_{\in}
      \left(      
         |\Psi_{z+}\rangle
         ,
         \mathcal{X}_{+}
      \right)
      \mkern-3mu
   \big)
   =
   0
   \;\;\;\;  ,
\end{equation}
\\[-35pt]

\begin{equation}  
   \Delta
   \mathrm{H}
   \left(
      {[\mkern-3.3mu[
         Z_{\pm}
      ]\mkern-3.3mu]}_v
   \right)
   =
   \mathrm{H}
   \big(
      \mathfrak{P}_{\in}
      \left(      
         |\Psi_{z-}\rangle
         ,
         \mathcal{Z}_{\pm}
      \right)
      \mkern-3mu
   \big)
   -
   \mathrm{H}
   \big(
      \mathfrak{P}_{\in}
      \left(      
         |\Psi_{z+}\rangle
         ,
         \mathcal{Z}_{\pm}
      \right)
      \mkern-3mu
   \big)
   =
   0
   \;\;\;\;  .
\end{equation}
\smallskip

\noindent In contrast, the process of the nondeterministic and nonreversible kind (that is, ``arbitrary change'' in von Neumann's formulation), $|\Psi_{z+}\rangle \mapsto |\Psi_{x+}\rangle$, corresponds to the changes in the valuational entropies\smallskip

\begin{equation}  
   \Delta
   \mathrm{H}
   \left(
      {[\mkern-3.3mu[
         X_{+}
      ]\mkern-3.3mu]}_v
   \right)
   =
   \mathrm{H}
   \big(
      \mathfrak{P}_{\in}
      \left(      
         |\Psi_{x+}\rangle
         ,
         \mathcal{X}_{+}
      \right)
      \mkern-3mu
   \big)
   -
   \mathrm{H}
   \big(
      \mathfrak{P}_{\in}
      \left(      
         |\Psi_{z+}\rangle
         ,
         \mathcal{X}_{+}
      \right)
      \mkern-3mu
   \big)
   =
   -
   \log_{\beta}2
   \;\;\;\;  ,
\end{equation}
\\[-35pt]

\begin{equation}  
   \Delta
   \mathrm{H}
   \left(
      {[\mkern-3.3mu[
         Z_{\pm}
      ]\mkern-3.3mu]}_v
   \right)
   =
   \mathrm{H}
   \big(
      \mathfrak{P}_{\in}
      \left(      
         |\Psi_{x+}\rangle
         ,
         \mathcal{Z}_{\pm}
      \right)
      \mkern-3mu
   \big)
   -
   \mathrm{H}
   \big(
      \mathfrak{P}_{\in}
      \left(      
         |\Psi_{z+}\rangle
         ,
         \mathcal{Z}_{\pm}
      \right)
      \mkern-3mu
   \big)
   =
   +
   \log_{\beta}2
   \;\;\;\;  ,
\end{equation}
\smallskip

\noindent which represent the acquisition of information about the spin of the qubit along the $x$-axis (and so the quantum state collapse from the full state $|\Psi_{z+}\rangle =\{|\Psi_{x\pm}\rangle\}$ to just one of the basis eigenstates, $|\Psi_{x+}\rangle$) together with the loss of the original information about the qubit's spin along the $z$-axis, in that order.\\

\noindent As the above analysis demonstrates, the presence of two separate processes of the state $|\Psi\rangle$ change, i.e., Schr{\"o}dinger's evolution and ``arbitrary change'', is not optional but rather proper to the Hilbert space formalism of quantum mechanics.\\

\noindent This means that it is not the case that the quantum measurement problem can be avoided by way of an interpretation that allows of only one process for changing the pure quantum state.\\

\bibliographystyle{References}

\end{document}